# The Reconfiguration Pattern of Individual Brain Metabolic Connectome for Parkinson's Disease Identification

Weikai Li[1,2,3,#], PhD; Yongxiang Tang[1,#], MD, PhD; Zhengxia Wang[4], PhD; Shuo Hu[1,5]*, MD, PhD; Xin Gao[3], PhD*;

**Affiliations:**

[1] Department of Nuclear Medicine (PET Center), XiangYa Hospital, Changsha, Hunan, P.R. China.

[2] College of Computer Science and Technology, Nanjing University of Aeronautics and Astronautics, Jiangsu, P.R. China.

[3] Shanghai Universal Medical Imaging Diagnostic Center, Shanghai, P.R. China.

[4] School of Computer Science and Cyberspace Security, Hainan University, Hainan, 570228, P. R. China.

[5] Key Laboratory of Biological Nanotechnology of National Health Commission, Xiangya Hospital, Central South University, Changsha, Hunan, P.R. China.

[#] **Co-first authors:** Weikai Li and Yongxiang Tang contributed equally to this work.

*Corresponding Author:

**Shuo Hu**, Department of Nuclear Medicine (PET Center), Key Laboratory of Biological Nanotechnology of National Health Commission, XiangYa Hospital, Central South University, 87 Xiangya Road, Changsha, Hunan 410008, P. R. China.

Tele: (+86)(731)-89753869; Fax: +86-0731-89753869

E-mail: hushuo2018@163.com

**Xin Gao**, Department of PET/MR, Universal Medical Imaging Diagnostic Center, Shanghai 20030, P.R. China

Tele: (+86) 15901613704

E-mail: george.ssmu@163.com

**ORCID**

Weikai Li https://orcid.org/0000-0002-5114-9660

Yongxiang Tang https://orcid.org/0000-0003-3987-8702

Shuo Hu https://orcid.org/0000-0003-0998-8943

**Abstract**

**Background:** Positron Emission Tomography (PET) with $^{18}$F-fluorodeoxyglucose ($^{18}$F-FDG) reveals metabolic abnormalities in Parkinson's disease (PD) at a systemic level. Previous metabolic connectome studies derived from groups of patients have failed to identify the individual neurophysiological details. We aim to establish an individual metabolic connectome method to characterize the aberrant connectivity patterns and topological alterations of the individual-level brain metabolic connectome and their diagnostic value in PD.

**Methods:** The $^{18}$F-FDG PET data of 49 PD patients and 49 healthy controls (HCs) were recruited. Each individual's metabolic brain network was ascertained using the proposed Jensen-Shannon Divergence Similarity Estimation (JSSE) method. The intergroup difference of the individual's metabolic brain network and its global and local graph metrics were analyzed to investigate the metabolic connectome's alterations. The identification of the PD from HC individuals was used by the multiple kernel support vector machine (MK-SVM) to combine the information from connection and topological metrics. The validation was conducted using the nest leave-one-out cross-validation strategy to confirm the performance of the methods.

**Results:** The proposed JSSE metabolic connectome method showed the most involved metabolic motor networks were PUT-PCG, THA-PCG, and SMA pathways in PD, which was similar to the typical group-level method, and yielded another detailed individual pathological connectivity in ACG-PCL, DCG-PHG and ACG pathways. The PD individuals showed higher nodal topological properties (including assortativity, modularity score, normalized characteristic path length, and characteristic path length) than HC individuals (all p < 0.05), whereas global efficiency and synchronization were lower in PD (both p < 0.05). Forty-five most significant connections were affected; most of the consensus connections in frontal, parietal, and occipital regions were decreased in PD patients while increased in the prefrontal, temporal, and subcortical regions. These aberrant functional network measures exhibited an ideal classification performance in the identifying of PD individuals from HC individuals at an accuracy of up to 91.84%.

**Conclusions:** The JSSE method identifies the individual-level metabolic connectome of $^{18}$F-FDG PET, providing more dimensional and systematic mechanism insights for PD. The proposed classification method also highlights the potential of an individual's connectome-based metrics to identify PD.

**Background**

With currently around 6.2 million people affected globally, Parkinson's disease (PD) is the second most common neurodegenerative movement disorder, and its prevalence is likely to increase over the coming decades [1]. Unfortunately, the diagnosis and disease-severity assessment of PD is mainly evaluated by clinical examination and follow-up. Typically, such disease is mainly characterized by motor symptoms (i.e., akinesia or bradykinesia, rigidity, and tremor) and non-motor symptoms such as cognitive impairment and psychiatric symptoms [2]. The motor symptoms related to PD are classically associated with the loss of dopamine-producing neurons in the substantia nigra [3, 4]. However, PD pathology also involves neurotransmitters other than dopamine and is distributed among many other cortical and sub-cortical brain regions participating in neural networks[5-7], making the clinical diagnosis and differentiating of early PD stages very challenging. The error rates for a clinical diagnosis of PD can be as high as 24%, even in specialized centers [8, 9].

Positron emission tomography with $^{18}$F-fluorodeoxyglucose ($^{18}$F-FDG PET) is a functional neuroimaging technique than provides a valuable imaging modality to identify and measure metabolic abnormalities in PD at a systemic level[10, 11]. Cerebral glucose metabolism, the primary energy source for neuronal activity, is closely associated with local neural function, integrity, and density. Previous studies have successfully provided considerable insight into the neurobiological basis, the relation between distinct brain regions, and associated clinical profiles in PD [12, 13]. The Parkinson's disease-related pattern (PDRP) of cerebral metabolism has been known since the early days of $^{18}$F-FDG PET imaging [14-16] and has already been considered as a biomarker for PD in clinical trials [17-21]. According to the updated standard clinical diagnostic criteria of PD [22, 23], the presence of normal presynaptic dopamine transporter (DAT) neuroimaging is one of the exclusions for ruling out PD. The DAT imaging is of limited value in differentiating PD from other atypical Parkinsonism because of its lack of specificity and whole-brain system-level analysis[24]. In particular, several $^{18}$F-FDG PET-based machine learning approaches have successfully promoted the proper application of metabolic PET imaging in PD [25, 26] and classified PD from healthy controls (HCs), which has achieved an empirical success [27-29]. However, such methods fail to consider the metabolic interactions of between-regions, potentially losing relevant information concerning individual differences in metabolic topology or network, and tended to be sensitive to the PET images' scanning parameters. This leads to instability of the biomarker, as the connectome tends to be more robust for identification as a second-order statistic[30].

Intuitively, the brain metabolic network detected in PD can help distinguish the disease from HCs and offering more evidence in support of parkinsonian pathophysiological mechanisms [31]. Thus, the individual metabolic network estimation can be clinically meaningful. An obvious goal is to reveal the robust brain metabolic network for PD analysis. Recently, a series of PET-based methodologies have been proposed to investigate metabolic networks [26, 32, 33]. However, most of the metabolic connectome studies have adopted group-level network methods for metabolic network modeling, which naturally sacrifice critical individual-level information. In this paper, inspired by the recent structural connectome studies using the distribution divergence to construct the individual network based on gray matter [34, 35], we opted to apply the Jensen-Shannon divergence similarity estimation (JSSE) to develop a new analytic methodology for individual-level metabolic brain network construction in $^{18}$F-FDG PET imaging and further provide an ensemble method towards PD understanding and identification.

Notably, several brain-network-based machine learning methods are proposed for PD analysis[36] or other neuro-diseases that[37, 38] that have achieved impressive results. These approaches mainly focus on the connection weight pattern, falling in discover the other connectome information (e.g., topology information). Specifically, inspired by the multi-view learning trick in the machine learning field [39], this study simultaneously utilized the information from the connection value, nodal graph measurement, and global graph measurement to accurately identify PD.

The present study's main purposes were to discover the altered pattern of the individual brain metabolic connectome, including its connection, nodal graph metrics, and global graph metrics. Subsequently, towards an accurate diagnosis of PDs from HCs, based on the multi-kernel trick, we identified (i) the most discriminative nodal features of the brain connectome and predominant brain regions in PD patients, (ii) achieved the accurate and automatic classification of PD patients and HCs, and (iii) analyzed the changing patterns in the brain network.

**Methods**

**Patients**

A total of 49 patients (33 male and 16 female participants, 53.94 ± 11.16 years), who were diagnosed with idiopathic PD based on the International Parkinson and Movement Disorder Society (MDS) diagnostic criteria [40] and scanned with $^{18}$F-FDG PET, were consecutively enrolled in this study between January 2018 and December 2019. We excluded patients with a history of head injury, cerebral stroke, intracranial operation, psychiatric illness, and substance use disorder. Detailed clinical participants' information can be found in **Table 1**. Also, a total of 49 HCs matched for similar age, education, and gender distribution were randomly recruited to obtain normative data. None of the HCs had a history of cognitive impairment, psychiatric illness, central nervous system disease, or head injury. All participants provided written informed consent following the Declaration of Helsinki. All aspects of the study were approved by the Studies Institutional Review Board Xiangya Hospital, Central South University.

**$^{18}$F-FDG PET Image acquisition and processing**

The $^{18}$F-FDG PET was performed at the PET Center of Xiangya Hospital using a Discovery Elite PET/computed tomography (CT) scanner (GE Healthcare, Boston, MA, USA). A CT transmission scan was performed for photon attenuation correction. Participants were injected intravenously with $^{18}$F-FDG (3.7 MBq/kg), required to rest for 45–60 min in the supine position on the PET scanner bed with their eyes closed and then underwent a 10 min scan using a 3-dimensional (3D) model. Images were reconstructed utilizing an ordered-subset expectation maximization algorithm with 6 iterations and 6 subsets method. The $^{18}$F-FDG PET imaging data processing was performed using the statistical parametric mapping (SPM) software (Wellcome Department of Cognitive Neurology, London, UK) implemented on MATLAB. Individual $^{18}$F-FDG PET image volumes were spatially normalized into standard stereotactic Montreal Neurological Institute (MNI) space with linear and nonlinear 3D transformations. To facilitate comparison across all participants, the intensity of images was globally normalized. After that, the automated anatomical labeling (AAL) [41]atlas was applied to segment the cerebral cortex into 90 regions (45 for each hemisphere without cerebellum).

**Individual JSSE metabolic network construction**

The distribution-divergence-based method has been successfully implemented for the individual morphological brain network construction [34, 35, 42]. However, so far, only a few studies have constructed individual metabolic networks from $^{18}$F-FDG PET imaging. This paper assumes that the $^{18}$F-FDG PET signal across brain regions indicates metabolic connections subserving inter-regional information transfer [43]. A relatively high resting signal-to-noise $^{18}$F-FDG PET signal in a region of interest (ROI) reflects the relative glucose metabolism rate. Thus, such a putative relationship offers a plausible approach to characterize inter-neuronal information transfer. Note that most of the existing works have constructed the network by the Kullback-Leibler (KL) [44] divergence (relative entropy):

$$D_{KL}(\mathbb{P}||\mathbb{Q}) = \int_{\mathcal{X}} \left( \mathbb{P}(x) \log \frac{\mathbb{P}(x)}{\mathbb{Q}(x)} + \mathbb{Q}(x) \log \frac{\mathbb{Q}(x)}{\mathbb{P}(x)} \right) dx \quad (1)$$

where $\mathbb{P}$ and $\mathbb{Q}$ represent the probability density functions (PDFs) of voxel intensities in a pair of ROIs; however, the KL-divergence is asymmetric. Thus, herein, we instead used the JSSE to capture the statistical relationship of the similarity of cerebral glucose metabolism in any 2 regions, which can then delineate individual metabolic connections. We represented brain nodes by 90 ROIs from the AAL atlas parcellation for depicting the individual metabolic network. Globally normalized $^{18}$F-FDG uptake in each ROI was used to generate a region × region correlation matrix (90 × 90) for each participant. The intensity of voxels in each ROI was extracted and used to estimate the PDF of the corresponding ROI by using kernel density estimation [45]. We then derived the metabolic connections as the Jensen-Shannon (JS) divergence (relative entropy) according to the mathematical equation:

$$D_{JS}(\mathbb{P}||\mathbb{Q}) = \frac{1}{2}[D_{KL}(\mathbb{P}||\mathbb{M}) + D_{KL}(\mathbb{Q}||\mathbb{M})] \quad (2)$$

where $\mathbb{M} = 0.5 \times (\mathbb{P} + \mathbb{Q})$ and $D_{KL}(\cdot\,|\,\cdot)$ are the KL divergence. Here, we used the JS divergence as a measure of metabolic connectivity to construct the adjacency matrix. In this way, this adjacency matrix describes pairwise metabolic connectivity, where the metabolic connection strength between region $i$ and $j$ can be represented by the corresponding element in this adjacency matrix.

**Computation of graph metrics**

We aimed to investigate the altered reconfiguration pattern of individual brain metabolic connectome for PD. Based on binary undirected matrices, we systematically analyzed the functional brain network's global and local properties with the Graph Theoretical Network Analysis Toolbox[1]. Specifically, the global metrics includes clustering coefficient ($C_p$), characteristic path length ($L_p$), normalized clustering coefficient ($\gamma$), normalized characteristic path length ($\lambda$), small-world ($\sigma$), global efficiency ($E_{global}$) and modularity score ($Q$) [46]. Also, the nodal property includes degree centrality, nodal efficiency, betweenness centrality, shortest path length and nodal clustering coefficient. The definition of these measurements can be found in the work of Wang et al [47]. Both global and nodal graph metrics were applied to characterize the different patterns of connections in the brain network, as shown in **Table 2**. Notably, we compared the network size at the different sparsity thresholds (of 0.02–0.5, with steps of 0.01), and a sum of 49 values of the corresponding node attributes under the sparsity threshold was obtained. We then took the sum of 49 values for each node (area under the curve, AUC) as input for the attributes to train the classifier, so there was only one value corresponding to one graph metric.

---

[1] www.nitrc.org/projects/gretna/

**Feature combination and PD identification**

Towards an accurate identification of the PDs from the HCs, we proposed to combine the information from the connection weights, nodal, and global graph metrics. Specifically, as an attempt, this paper adopted the kernel combination trick for information combination and utilized the multi-kernel support vector machine (MK-SVM) for PD identification. Specifically, the MK-SVM method in this study can be conducted as follows. In particular, suppose that there are $n$ training samples with connections values and graph metrics, let $x_i^1$, $x_i^2$, and $x_i^3$ represent the connection weight, the graph metrics, and nodal graph metrics of the $i$-th sample, respectively. Denoting that $y_i \in \{1, -1\}$ be the corresponding label; the MK-SVM solves the following primal problem:

$$\min_W \frac{1}{2} \sum_{m=1}^{3} \beta_m \|w^m\|^2 + C \sum_{i=1}^{n} \xi_i$$

$$\text{s.t.} \ y_i \left( \sum_{m=1}^{3} \beta_m (w^m)^T \phi^m(x_i^m) + b \right) \geq 1 - \xi_i \quad (3)$$

$$\xi_i \geq 0, i = 1, 2, \ldots, n$$

where $\phi^m$ represents the transform from the original space in $m$-th data to the Represent Hilbert Kernel Space (RHKS), $w^m$ represents the hyperplane in RHKS, and $\beta_m$ denotes the corresponding combining weight on the $m$-th attribute. Next, the dual form of MK-SVM can be represented as:

$$\max_\alpha \sum_{i=1}^{n} \alpha_i - \frac{1}{2} \sum_{i,j} \alpha_i \alpha_j y_i y_j \sum_{m=1}^{3} \beta_m k^m(x_i^m, x_j^m)$$

$$\text{s.t.} \ \sum_{i=1}^{n} \alpha_i y_i = 0 \quad (4)$$

$$0 \leq \alpha_i \leq C, i = 1, 2, \ldots n$$

where $k^m(x_i^m, x_j^m) = \phi^m(x_i^m)^T \phi^m(x_j^m)$ and is the kernel matrix on the $m$-th data. After we trained the model, we tested the new samples $x = \{x_1, x_2, \cdots, x_M\}$. The kernel between the new test sample and $i$-th training sample on the $m$-th modality is defined as $k^m(x_i^m, x^m) = \phi^m(x_i^m)^T \phi^m(x^m)$. In the end, the predictive level based on MK-SVM can be formulated as follows:

$$f(x_1, x_2, \ldots, x_M) = \text{sign}(\sum_{i=1}^{n} y_i \alpha_i \sum_{m=1}^{M} \beta_m k^m(x_i^m, x^m) + b) \tag{5}$$

To illustrate the performance gain of the information combination from different views, such as connection and metrics, we employed the most commonly used and the simplest linear kernel as $k^m(x_i^m, x_j^m)$, which is given as follows:

$$k^m(x_i^m, x_j^m) = x_i^{mT} x_j^m \tag{6}$$

**Feature selection and validation**

To confirm the effectiveness of the proposed PD identification, we conducted the strictest nest leave-one-outcross-validation (LOOCV) strategy to verify the performance of the methods due to the small sample size, in which only one subject was left out for testing while the others are used to train the models and obtain the optimal parameters. For the choice of optimal parameters, an inner LOOCV was conducted on the training data using a grid-search strategy. The range of the hyper-parameter $C$ was $2^{-5}$ to $2^5$. Meanwhile, to alleviate the interference from the feature selection procedure, we selected the simplest feature selection method (t-test with $p < 0.05$) to select the nodal graph metric and the connection weight in our experiment [38]. All data-processing and classification procedures used in our study are shown in **Figure 1**.

**Results**

**Global graph metrics of metabolic brain connectome**

The global graph metrics of PD and HC groups are shown in **Table 3**. With $Ar$, $Q$, $Hr$, $E_{local}$, $C_p$, $\gamma$, $\lambda$, and $L_p$ increased, whereas $E_{global}$, $\sigma$, and $S_r$ decreased in the PD groups. Statistical analyses revealed that the $A_r$, $Q$, $\lambda$, and $L_p$ of PD patients were significantly higher than those in the HC group, illustrating the significantly modularity change in the PD patients. Besides, the $S_r$ and $E_{global}$ were significantly lower in the PD groups compared with the HC groups ($p < 0.05$).

**Degree analysis the metabolic brain connectome**

To investigate the degree distribution of the estimated metabolic brain connectome, we visualized each node's mean degree of each node in the PD and HC groups. As shown in **Figure 2**, the degree in frontal and parietal regions tends to be decreased in PD while tending to be increased in the prefrontal and subcortical regions. Specifically, the 19 significant nodes with the average degree in the PD and HC groups are listed in **Table 4**.

The nodes with a degree of standard deviations higher than the mean of the degree of all nodes were identified as degree hub nodes [48]. Here, according to the definition of " hubs, " we identified hub nodes in PD patients and HCs, separately, in **Figures 3 and 4**. By comparing the hub nodes between 2 groups in the same modal network, it was obvious found that most of them were overlapped. Also, it is worth noting that several specific hub nodes existed that corresponded to different groups. For instance, in a metabolic network based on PET, the PCUN.L, SOG.R, and IOG.R only appeared in the hub nodes of the HCs, while the MTG.R and PCUN.R only appeared in the PD group as the hub node.

**Between analysis the metabolic brain connectome**

Similarly, to investigate the betweenness distribution of the estimated metabolic brain connectome, we visualized the mean betweenness of the PD and HC groups in **Figure 5**. The result illustrates that betweenness in frontal and parietal regions tends to be decreased in PD while tending to be increased in the temporal and subcortical regions. The 15 significant nodes with the average betweenness in PD and HC groups are listed in **Table 5**.

We also identified betweenness hub nodes of PD patients and HCs in **Figures 6 and 7**, respectively. By comparing the betweenness hub nodes between the PD group and HC groups in the same modal network, several specific hub nodes correspond to different groups. The INS.R/L, ORBsup.L, SMG.R/L, ORBmid.R/L, and PCUN.R only appeared in the hub nodes of the HCs, while the IOG.L, SFGmed.R and ACG.L only appeared in the PD group as the hub node.

**Classification results**

To evaluate the classification performance of the information combination methods and the proposed JSSE, we conducted several quantitative measurements, including accuracy, sensitivity and specificity to evaluate the classification performance. The mathematical definitions of these 3 measures were given as follows:

$$Accuracy = \frac{TruePostive + TrueNegative}{TruePostive + FalsePostive + TrueNegative + FalseNegative}, \quad (7)$$

$$Sensitivity = \frac{TruePostive}{TruePostive + FalseNegative}, \quad (8)$$

$$Specificity = \frac{TrueNegative}{TrueNegative + FalsePostive}, \quad (9)$$

*TruePositive* is the number of the positive participants that are correctly classified in the PD identification task. Similarly, *TrueNegative*, *FalsePostive* and *FalseNegative* are the numbers of their corresponding subjects, respectively. Also, the receiver operating characteristic curve (ROC) and AUC of these methods is also provided.

To validate the information combination trick, we also reported the single kernel SVM classification result based on the connection, global metrics and nodal metrics. The results are shown in **Table 6**. The ROC curve result is shown in **Figure 8**, where the performances of information combination results are superior to that of single kernel methods, indicating the proposed methods' rationality. Additionally, the C+G+N method (combines the information of Connection, Global metrics, and Nodal Metrics) achieves the outperforming result in all 4 measurements, demonstrating its effectiveness. Also, with DeLong's non-parametric statistical significance test [49], the proposed C+G+N methods are significantly superior to Connection, Global, and Nodal under 95% confidence interval (CI) with p-values of 0.0482, $4 \times 10^{-6}$ and 0.0115, respectively. The superior performance illustrated that the information combination scheme could effectively improve the classification performance.

**Consensus significant metabolic connections**

As mentioned above, we selected the consensus connections with p < 0.05 in each loop. A total of 45 consensus connections are shown in **Figure 9**. Most of the consensus connections in frontal, prefrontal, and occipital regions were decreased in PD patients while increasing in the parietal and subcortical regions. There were 35 nodes with consensus connections, and the degree of the consensus connection is listed in **Table 7**.

**Discussion**

We have presented a new analytic methodology for individual-level metabolic brain network construction in $^{18}$F-FDG PET imaging and its graph theory metrics to investigate the alterations of PD's metabolic connectome. The findings of this study provide insight into the compensatory mechanism of underlying PD. The proposed classification method highlights the potential of connectome-based metrics for the identification of PD. Although $^{18}$F-FDG PET based on metabolic PDRP and group-level network methods has been shown to achieve a great success for PD analysis. such approaches cannot identify individual neurophysiological details and the metabolic perturbations basal ganglia and cortical connectivity that underlie the cardinal motor features of PD. The ascendancy of our novel JSSE approach may afford individual neurophysiological mechanisms of a multitude of non-motor features.

The PDRP was first identified by Eidelberg et al. [50, 51], and several studies have made the diagnosis of PD with the assistance of metabolic PDRP markers [10, 52]. The metabolic network is highly reproducible and specific for PD, which may help distinguish PD from HC [53]. During group-level analyses, the clinical and pathophysiologic correlates of the PDRP have been investigated extensively [10], which has potentially sacrificed or obscured salient individual differences within a group. Specifically, we used the JSSE to directly estimate the symmetric metabolic network based on the $^{18}$F-FDG PET. Our individualized metabolic JSSE network analysis revealed subtle deviations in connection value, nodal graph measurement, and global graph measurement that were powerfully predictive of PD from HC. Our findings of decreased connectivity patterns in the frontal, parietal, and occipital regions and increased in the prefrontal, temporal, and subcortical regions, recapitulated the results of previous $^{18}$F-FDG PET group-based comparisons [10, 12, 31, 54]. Overall, based on a calculation of relative entropy, we found that applying JSSE to $^{18}$F-FDG PET data is a compelling approach for revealing an individual's metabolic connectivity networks and novel insights into the nature of metabolic disturbances in PD clinical characteristics, which could offer a new perspective of their application in future studies of PD.

Because our JSSE brain connectome approach can measure global and local graph metrics of the metabolic network, it could powerfully identify PD's salient properties. Our main finding was that assortativity coefficient, modularity score, normalized characteristic path length, and characteristic path length of PD patients were higher than that of HC group, whereas synchronization and global efficiency were lower in the PD groups. The degree and between analysis of metabolic brain connectome appeared to change in our comparison of HC and AD groups. The main connectome finding was that PD entails disruption of modules (or subnetworks) of the global network architecture and a loss of connectivity between those modules. These results are in accord with previous metabolic network studies showing analogous network changes in PD [26, 30, 51, 55].

It has also been shown that motor networks described under normal conditions are disturbed in PD. In our analysis from consensus significant metabolic connections, the most involved metabolic network was the PUT-PCG pathway. The PUT is an important part of the basal ganglia (BG), which is affected as a key node of the metabolic network and seems to be associated with impaired motor symptoms, which elegantly posited an explanation for rigidity and bradykinesia [56, 57]. Beyond the PUT-PCG pathway, THA-PCG and SMA were also shown to be involved. The SMA, a key region associated with motor symptoms, is also affected as a central node. These findings could be supported by studies showing the differential involvement of BG and cortex-striatum-thalamus-cortex and that both structures could modulate each other at the subcortical level [55, 58], which is similar to the typical group-level method. The dopaminergic system projects to motor pathways and the lateral prefrontal cortices via the caudate.

This frontostriatal system is intimately connected with the posterior parietal cortex and is related to executive and memory functioning. It could also be seen from our results that the projection areas downstream of individual pathological connectivity in the ACG-PCL, DCG-PHG, and ACG pathways were significantly affected. Therefore, the akinesia-rigidity-related metabolic connectivity networks validated in our study might help understanding certain circuits' pathophysiology and elucidating their specific roles in causing clinical symptoms. Besides these regions, our results showed that more connections (45 significant connections in total) of PD patients were affected, worthy of further study on a larger scale combining with clinical data and stratification.

As shown in our work, the brain's metabolic connectivity networks that may be affected by tremor symptoms associated with the cerebellum were not significantly affected. Previous studies have shown that the PDRP was not related to tremor, probably reflecting the distinct pathophysiologic origins of bradykinesia/rigidity and tremor [12, 59]. The tremor-related regions identified in previous studies were identical to the regions reported in the "dimmer switch" theory, in which the striato-pallidal circuit triggers tremor episodes (light switch). In contrast, the cerebello-thalamo-cortical circuit produces the tremor and controls its amplitude (light dimmer) [54, 59]. In the future, we will include the metabolic network related to the cerebellum in analysis, which may provide a better entry point for more clinical symptoms of PD, deeper mechanism research, and provide better diagnostic performance for PD diagnostic imaging markers.

Our present brain connectome approach could powerfully identify PD from HC because it can measure local network properties and the entire network. In this regard, our main finding was that PD entails disruption of modules (or subnetworks) of the global network architecture and a loss of connectivity between those modules. Several previous PET studies have likewise revealed an early failure of brain modules concerning the onset of cognitive dysfunction. In the present work, we observed that the metabolic connections could provide the effective biomarkers for identifying PD. Further, the topological information can also provide more discriminative information, since the MK-SVM, which combines all of this information, provides a more accurate diagnostic result.

There were several limitations to the current study that must be considered when interpreting our findings. Our data were collected retrospectively and derived from a limited number of patients. Other preoperative neuroimaging data and results of more sophisticated diagnostic tests such as structural magnetic resonance imaging (MRI) were not analyzed in this study, and the JSSE method was implemented for $^{18}$F-FDG PET images without partial volume effect (PVE) correction. Furthermore, our interpretation of a link between a metabolic network failure and PD pathology remains confirmed by DAT neuroimaging and clinical manifestations. Therefore, future studies should incorporate data other than glucose metabolism employed in the current study to verify the metabolic connectome method's applicability fully.

**Conclusions**

This study presents an advanced connectome analysis of 18F-FDG PET images based on a novel application of JSSE entropy measures, which had not been previously applied to the task of metabolic connectome analysis. This method sheds new light on the network abnormality underlying PD. Importantly, we have presented a novel understanding and more evidence to further path mechanism research in PD. Further, our finding demonstrates that information combinations from different views could achieve an ideal performance for identifying the PDs from NCs.

**Abbreviations**

3D: 3-dimensional; AAL: Automated anatomical labeling; Ar: Assortativity; AUC: Area under the curve; BG: Basal ganglia; CI: Confidence interval; Cp: Clustering coefficient; CT: Computed tomography; DAT: Dopamine transporter; E $_{global}$: Global efficiency; E $_{local}$: Local efficiency; $^{18}$F-FDG: $^{18}$F-fluorodeoxyglucose; HC: Healthy control; Hr: Hierarchy; JSSE: Jensen-Shannon Divergence Similarity Estimation; KL: Kullback-Leibler; LOOCV: Leave-one-outcross-validation; Lp: Characteristic path length; MDS: Movement disorder society MK-SVM: Multiple kernel support vector machine; MNI: Montreal Neurological Institute; MRI: Magnetic resonance imaging; PD: Parkinson's disease; PDF: Probability density function; PDRP: Parkinson's disease-related pattern; PET: Positron emission tomography; PVE: Partial volume effect; Q: Modularity score; RHKS: Represent Hilbert Kernel Space; ROC: Receiver operating characteristic curve; ROI: Region of interest; SPM: Statistical parametric mapping; Sr: Synchronization; γ: Normalized clustering coefficient; λ: Normalized characteristic path length; σ: Small-world

**Ethics approval and consent to participate**

Approval was obtained from the ethics committee of Ethical Commission of Medical Research Involving Human Subjects at Region of Xiangya Hospital, Central South University, China. The procedures used in this study adhere to the tenets of the Declaration of Helsinki. Informed consent was obtained from all individual participants or legal guardians included in the study.

**Consent for publication**

Not applicable

**Availability of data and materials**

The datasets and custom code generated during and/or analyzed during the current study are available from the corresponding author on reasonable request.

**Competing interests**

The authors declare that they have no competing interests.

**Funding**

This study was funded by grant no. 81801740 from National Natural Science Foundation of China, grant no. 2020JJ5922 from Natural Science Foundation of Hunan Province, grant no. 2019-RGZN-01079 from Shanghai Municipal Commission of Economy and Informatization, grant no. cstc2018jcyjA0398 from Natural Science Foundation Project of CQCSTC; and grant no. UV2020Z02 from Scientific Research Subjects of Shanghai Universal Medical Imaging Technology Limited Company

**Authors' contributions**

All authors contributed to the study conception and design. Material preparation, data collection and analysis were performed by Weikai Li, Yongxiang Tang, Zhengxia Wang, Shuo Hu and Xin Gao. The first draft of the manuscript was written by Weikai Li and Yongxiang Tang. Shuo Hu and Xin Gao commented on previous versions of the manuscript. All authors read and approved the final manuscript.

**Acknowledgments**

We extend our thanks to the XiangYa Hospital Neurology Department staff for their expert assistance in this work, SPM12 (Welcome Department of Cognitive Neurology, London, UK) and ITK-SNAP (www.itksnap.org) for freely providing software analysis tools.

**Figure legends**

**Figure 1. Data-processing and classification procedures employed in our study.** The JSSE was adopted to construct the metabolic network. Then, the nodal and global graph metrics were computed. In the end, the MK-SVM was adopted to combine these information for PD identification.

**Figure 2. Degree distribution of PD and NC groups.** The degree in frontal and parietal regions tends to be decreased in PD while tending to be increased in the prefrontal and subcortical regions.

**Figure 3. Degree hub nodes in PD groups.**

**Figure 4. Degree hub nodes in HC groups**.

**Figure 5. Betweenness distribution of PD and HC groups.** The betweenness in frontal and parietal regions tends to be decreased in PD while tending to be increased in the temporal and subcortical regions.

**Figure 6. Betweenness hub nodes in PD groups.**

**Figure 7. Betweenness hub nodes in HC groups.**

**Figure 8. The ROC results of different methods.**

**Figure 9. The most consensus connections.** The most consensus connections mapped on the International Consortium for Brain Mapping (ICBM) 152 template using the BrainNet Viewer software package http://nitrc.org/projects/bnv/ and circularGraph, shared by Paul Kassebaumb http://www.mathworks.com/matlabcentral/fileexchange/48576-circulargraph). (**Left**) The arc thickness indicates the discriminative power of an edge, which is inversely proportional to the estimated p-values. (**Right**) The connectivity matrices of the fully connected network of PD compared to NCs are shown. The 45 most significant connections were retained, with green and red lines representing connection weights that are decreased and increased in PDs, respectively.

**Tables legends**

**Table 1. Demographic and clinical characteristics in PD patients and HCs**

**Table 2. Selected global and local graph metrics**

**Table 3. Global and local graph metrics of the metabolic brain connectome.**

**Table 4. 19 significant nodes with the average degree in PD and NC group.**

**Table 5. 15 significant nodes with the average betweenness in PD and NC group**

**Table 6. Classification performance corresponding to different methods.**

**Table 7. Degree of the consensus connection**

**Figures**

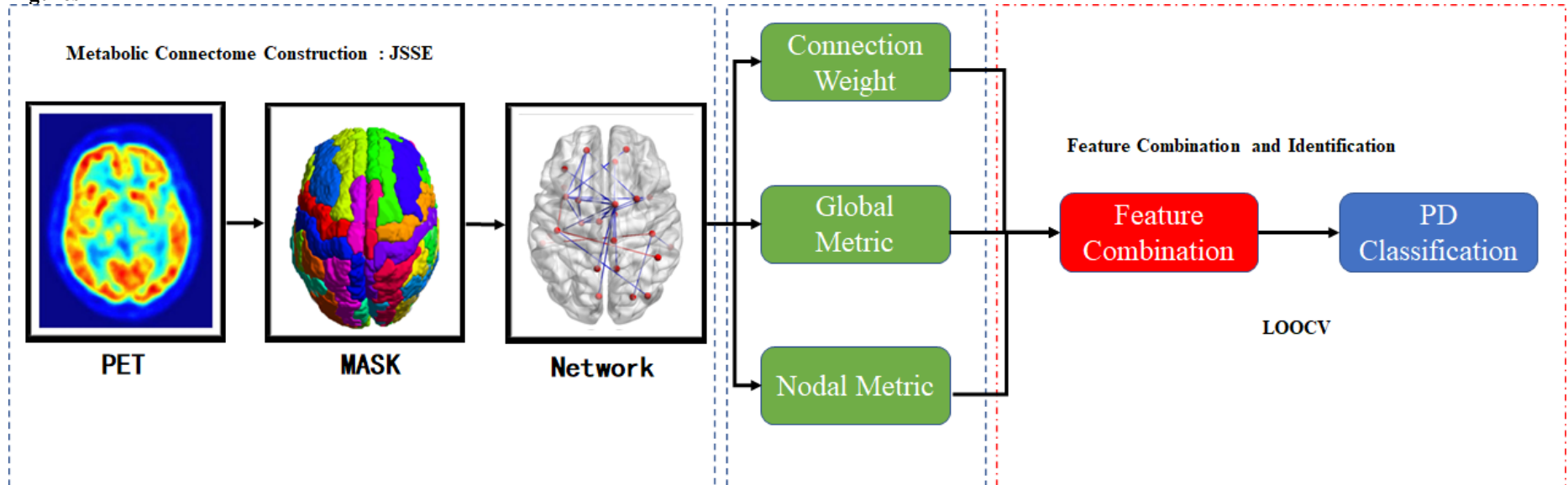

**Figure 1.** Data-processing and classification procedures employed in our study.
The JSSE was adopted to construct the metabolic network. Then, the nodal and global graph metrics were computed. In the end, the MK-SVM was adopted to combine these information for PD identification.

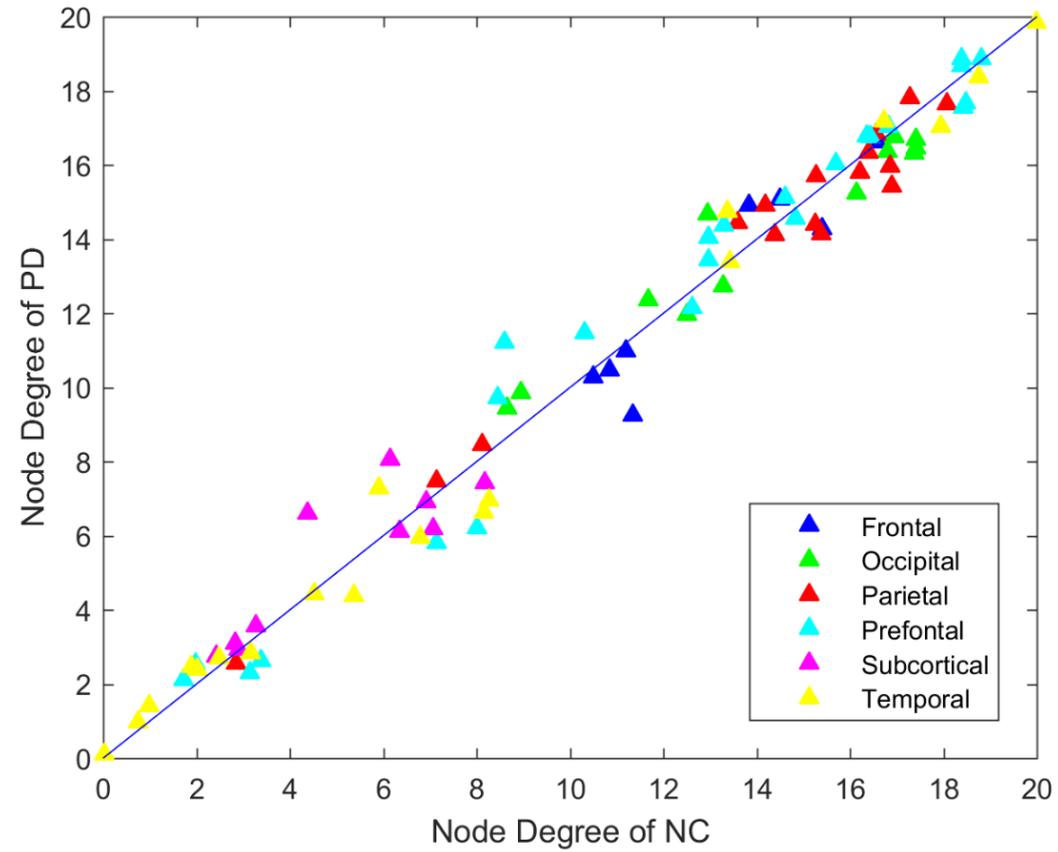

**Figure 2.** Degree distribution of PD and NC groups.
The degree in frontal and parietal regions tends to be decreased in PD while tending to be increased in the prefrontal and subcortical regions.

**Figure 3.** Degree hub nodes in PD groups.

**Figure 4.** Degree hub nodes in HC groups.

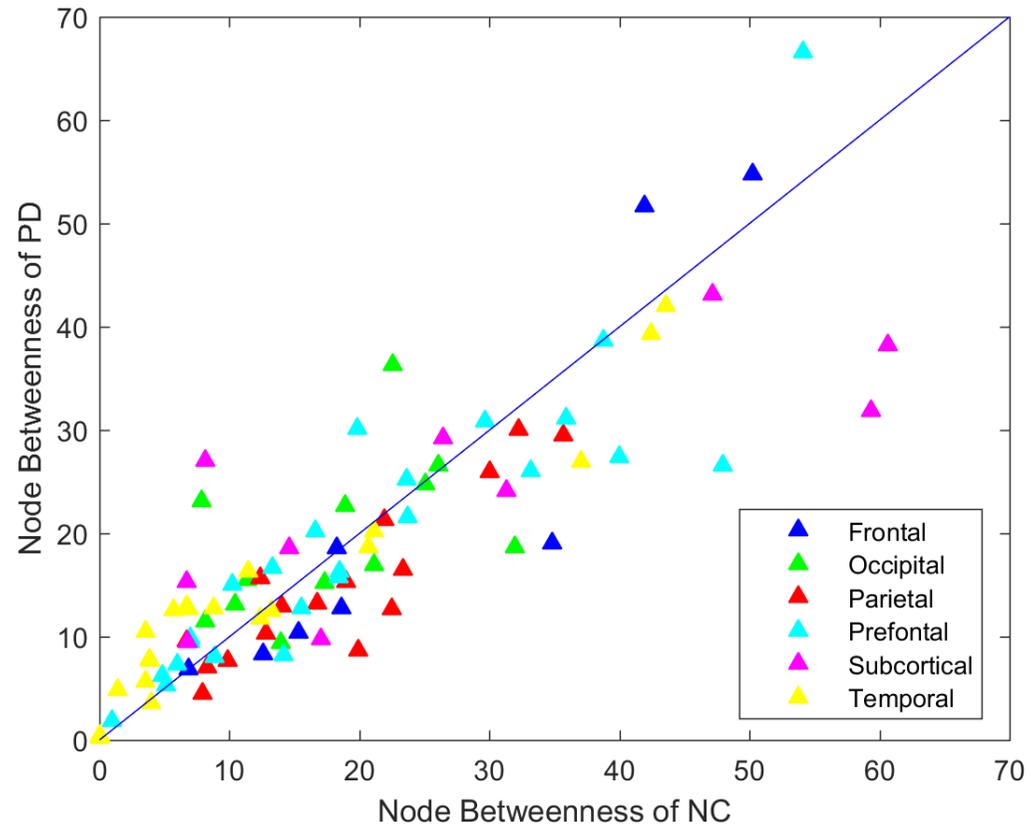

**Figure 5.** Betweenness distribution of PD and HC groups.
The betweenness in frontal and parietal regions tends to be decreased in PD while tending to be increased in the temporal and subcortical regions.

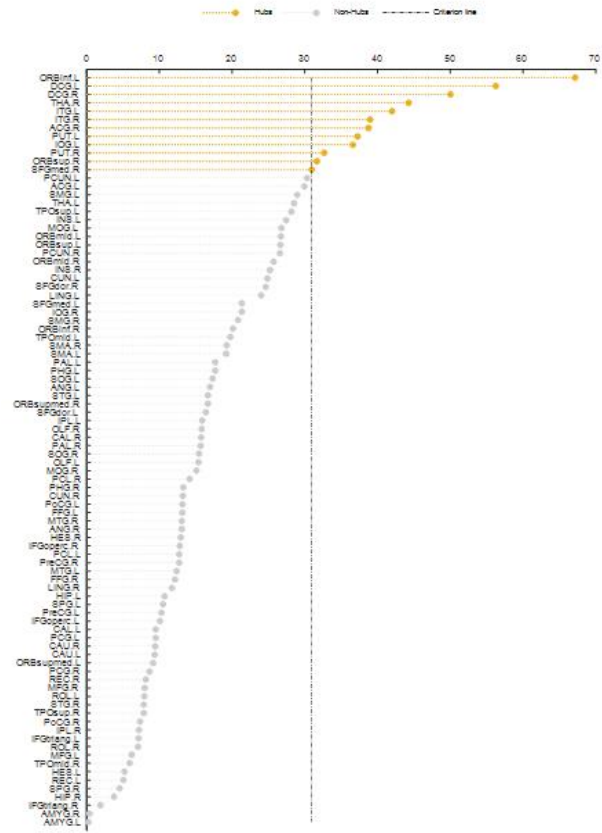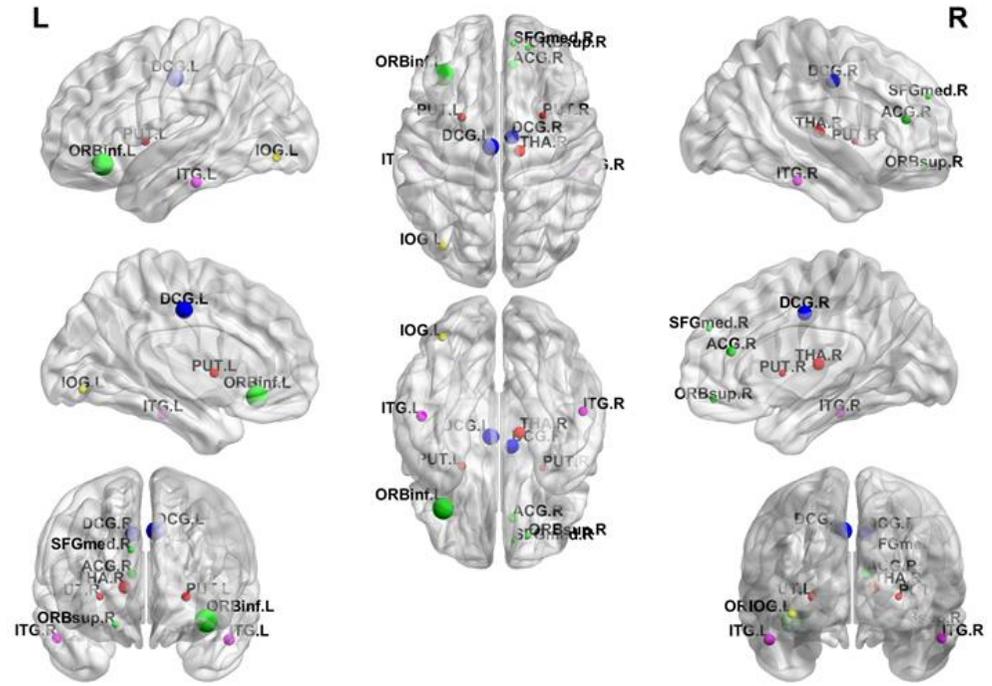

**Figure 6.** Betweenness hub nodes in PD groups.

**Figure 7.** Betweenness hub nodes in HC groups.

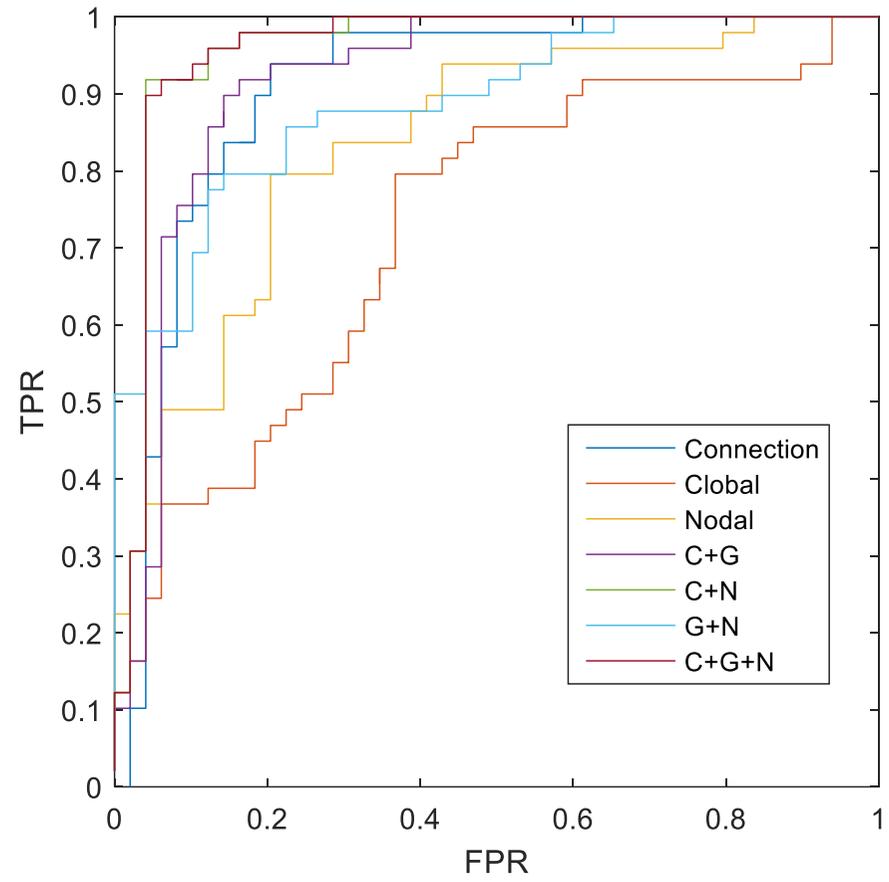

**Figure 8.** The ROC results of different methods.

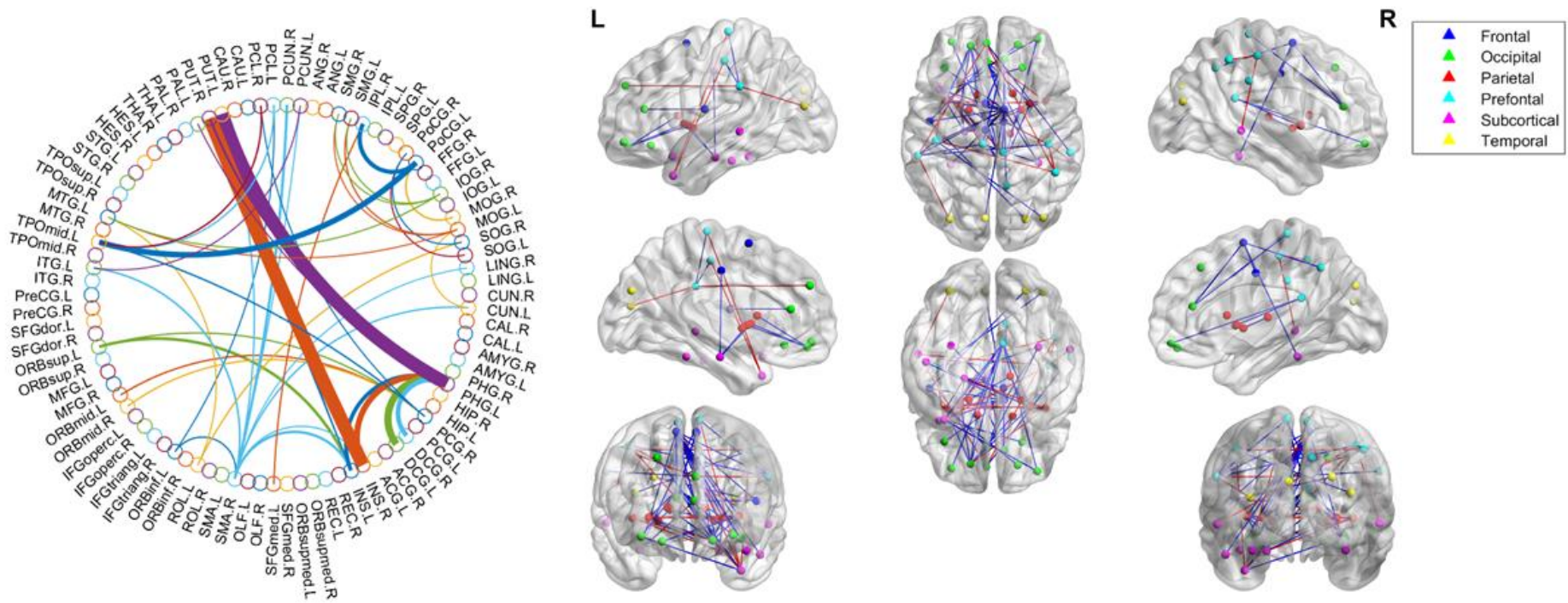

**Figure 9.** The most consensus connections

The most consensus connections mapped on the International Consortium for Brain Mapping (ICBM) 152 template using the BrainNet Viewer software package http://nitrc.org/projects/bnv/ and circularGraph, shared by Paul Kassebaumb http://www.mathworks.com/matlabcentral/fileexchange/48576-circulargraph). (**Left**) The arc thickness indicates the discriminative power of an edge, which is inversely proportional to the estimated p-values. (**Right**) The connectivity matrices of the fully connected network of PD compared to NCs are shown. The 45 most significant connections were retained, with green and red lines representing connection weights that are decreased and increased in PDs, respectively.

**Tables**

**Table 1. Demographic and clinical characteristics in PD patients and HCs**

| Variable | Controls (n = 49) | PD (n = 49) | p |
|---|---|---|---|
| Age (years) | 52.12±9.84 | 53.94±11.16 | 0.832 |
| Sex (male/female) | 30/19 | 33/16 | 0.577 |
| Education (years) | 13.44±3.15 | 12.37±4.06 | 0.163 |
| Disease duration (months) | NA | 60.2 | NA |
| UPDRS-III score | NA | 23.2 | NA |

NC, normal control; PD, Parkinson's disease; UPDRS, unified Parkinson's disease rating scale.

**Table 2. Selected global and local graph metrics**

| Global graph metrics | Local graph metrics |
|---|---|
| Clustering coefficient ($C_p$) | Degree centrality |
| Characteristic path length ($L_p$) | Nodal efficiency |
| Normalized clustering coefficient ($\gamma$) | Betweenness centrality |
| Normalized characteristic path length ($\lambda$) | Shortest path length |
| Small-world ($\sigma$) | Nodal clustering coefficient |
| Global efficiency ($E_{global}$) | |
| Local efficiency ($E_{local}$) | |
| Modularity score ($Q$) | |
| Assortativity ($Ar$) | |
| Hierarchy ($Hr$) | |
| Synchronization ($Sr$) | |

**Table 3. Global and local graph metrics of the metabolic brain connectome.**

| Global graph metrics | PD | NC |
|:---:|:---:|:---:|
| $Ar^*$ | 0.1502±0.02 | 0.1432±0.02 |
| $Q^*$ | 10.8503±1.25 | 10.4082±0.93 |
| $Hr$ | 0.0080±0.02 | 0.0068±0.01 |
| $E^*_{global}$ | 0.2231±0.01 | 0.2274±0.00 |
| $E_{local}$ | 0.3351±0.01 | 0.3339±0.01 |
| $C_p$ | 0.2881±0.01 | 0.2862±0.01 |
| $\gamma$ | 0.6408±0.05 | 0.6261±0.04 |
| $\lambda^*$ | 0.5275±0.01 | 0.5210±0.01 |
| $\sigma$ | 0.5401±0.04 | 0.543±0.04 |
| $L_p^*$ | 1.0529±0.05 | 1.0374±0.04 |
| $Sr^*$ | -0.6865±0.89 | -0.1746±0.28 |

$A_r$, assortativity; $C_p$, clustering coefficient; $E_{global}$, global efficiency; $E_{local}$, local efficiency; $Hr$, hierarchy; $L_p$, characteristic path length; NC, normal control; PD, Parkinson's disease; Q, modularity score; $S_r$, synchronization; γ, normalized clustering coefficient; λ, normalized characteristic path length; σ, small-world. * p-value < 0.05.

**Table 4. 19 significant nodes with the average degree in PD and NC group.**

|  | MFG.L | ROL.L | SMA.L | OLF.L | OLF.R | ORBsupmed.L | ORBsupmed.R | ACG.R | AMYG.L | AMYG.R |
|---|---|---|---|---|---|---|---|---|---|---|
| NC | 12.85745 | 11.44102 | 13.74214 | 1.741122 | 1.624388 | 9.476735 | 9.89449 | 7.774898 | 0.02102 | 0.025306 |
| PD | 13.95122 | 9.117347 | 15.23551 | 2.537143 | 2.171122 | 11.49398 | 11.73776 | 6.24102 | 0.105306 | 0.108571 |
| $P_{value}$ | 0.031289 | 0.002185 | 0.019944 | 0.011798 | 0.044952 | 0.02319 | 0.036806 | 0.016161 | 0.003493 | 0.004302 |

|  | FFG.R | PoCG.R | SMG.L | THA.L | THA.R | HES.L | STG.R | MTG.L | TPOmid.L |
|---|---|---|---|---|---|---|---|---|---|
| NC | 7.918469 | 15.37347 | 17.04286 | 4.593061 | 6.258469 | 1.594082 | 13.28173 | 18.08561 | 8.450306 |
| PD | 6.584082 | 14.055 | 15.35806 | 6.773265 | 8.104592 | 2.565816 | 14.81122 | 16.95786 | 6.572143 |
| $P_{value}$ | 0.021963 | 0.026412 | 0.021154 | 0.000653 | 0.005116 | 0.033751 | 0.014103 | 0.012137 | 0.023848 |

NC, normal control; PD, Parkinson's disease

**Table 5. 15 significant nodes with the average betweenness in PD and NC group**

|  | IFGtriang.R | ROL.L | SMA.R | OLF.R | SFGmed.R | HIP.L | SOG.L | MTG.L |
|---|---|---|---|---|---|---|---|---|
| NC | 1.165445 | 14.3233 | 29.50238 | 8.682077 | 23.48307 | 5.262653 | 24.10385 | 16.86597 |
| PD | 1.881727 | 7.91026 | 19.23082 | 15.84762 | 30.91199 | 10.74993 | 17.31138 | 12.40085 |
| $P_{value}$ | 0.049115 | 0.000901 | 0.004348 | 0.016143 | 0.009278 | 0.012116 | 0.023121 | 0.011204 |
|  | SOG.R | PCL.L | CAU.R | PUT.L | THA.L | STG.R | TPOsup.R |  |
| NC | 20.12192 | 18.03614 | 13.97596 | 51.8123 | 12.30414 | 4.480052 | 4.366766 |  |
| PD | 15.48497 | 12.75296 | 9.408317 | 37.28866 | 28.51788 | 7.879975 | 7.879856 |  |
| $P_{value}$ | 0.048746 | 0.02918 | 0.046454 | 0.024638 | 0.001588 | 0.001181 | 0.041972 |  |

NC, normal control; PD, Parkinson's disease

**Table 6. Classification performance corresponding to different methods.**

| Method | Accuracy | Sensitivity | Specificity | AUC |
|---|---|---|---|---|
| Connection (C) | 87.76 | 79.59 | 85.71 | 0.9058 |
| Global (G) | 65.31 | 63.27 | 67.35 | 0.7242 |
| Nodal (N) | 76.53 | 81.63 | 71.43 | 0.8354 |
| C+G | 87.76 | 89.80 | 85.71 | 0.9171 |
| C+N | 90.82 | 93.88 | 87.76 | 0.9566 |
| G+N | 81.63 | 77.55 | 85.71 | 0.8879 |
| C+G+N | **91.84** | **93.88** | **89.90** | **0.9571** |

Note: C+G+N methods are significantly superior to Connection, Global, and Nodal under 95% confidence interval with p-value equals to 0.0482, $4 \times 10^{-6}$ and 0.0115 respectively.

**Table 7. Degree of the consensus connection**

| ROI | SMA.R | PHG.L | INS.L | PCG.R | MOG.L | PUT.L | TPOmid.L |
|---|---|---|---|---|---|---|---|
| Degree | 9 | 7 | 5 | 5 | 4 | 4 | 4 |
| ROI | PCL.R | ITG.L | IOG.R | PoCG.L | PUT.R | MTG.L | INS.R |
| Degree | 2 | 2 | 3 | 3 | 3 | 3 | 4 |
| ROI | SOG.R | PoCG.R | IPL.R | SMG.L | SMG.R | PCUN.L | PCL.L |
| Degree | 2 | 2 | 2 | 2 | 2 | 2 | 2 |
| ROI | ORBsup.L | ORBinf.L | ROL.L | CUN.L | MOG.R | PCUN.R | PAL.L |
| Degree | 2 | 2 | 2 | 2 | 1 | 1 | 1 |
| ROI | ORBmid.L | ORBmid.R | SFGmed.L | DCG.L | DCG.R | LING.R | SOG.L |
| Degree | 1 | 1 | 1 | 1 | 1 | 1 | 1 |